\documentclass[reprint,amsmath,amssymb, aps,pre,showpacs]{revtex4-1}

\usepackage{graphicx}
\usepackage{dcolumn}
\usepackage{bm}

\begin{document}

\title{Aggregation-fragmentation model of robust concentration gradient formation}

\author{Timothy E. Saunders}
\email{dbsste@nus.edu.sg}
\affiliation{
Mechanobiology Institute, National University of Singapore, Singapore \\
Department of Biological Sciences, National University of Singapore, Singapore \\
Institute of Molecular and Cell Biology, Proteos, Singapore
}

\date{\today}

\begin{abstract}
Concentration gradients of signaling molecules are essential for patterning during development and they have been observed in both unicellular and multicellular systems.  In subcellular systems, clustering of the signaling molecule has been observed. We develop a theoretical model of cluster-mediated concentration gradient formation based on the Becker-D{\"o}ring equations of aggregation-fragmentation processes.  We show that such a mechanism produces robust concentration gradients on realistic time and spatial scales so long as the process of clustering does not significantly stabilize the signaling molecule.  Finally, we demonstrate that such a model is applicable to the pom1p subcellular gradient in fission yeast. 
\end{abstract}


\maketitle

\section{\label{sec:intro}Introduction}

Individual cells make precise cell-fate decisions based on information from signaling networks during development.  A central paradigm of information transfer during development is the morphogen gradient, a spatially varying concentration profile \cite{Wolpert:1969wu}. The dynamics of morphogen gradient formation \cite{Muller:2013it} and interpretation \cite{Rogers:2011eh} has been the subject of intense study, both experimentally \cite{Kicheva:2007bh, Gregor:2007ce,  Balaskas:2012cf, HaskelIttah:2012ft, Warmflash:2012hk} and theoretically \cite{Eldar:2003eh, Bollenbach:2005ky,England:2005gm,Coppey:2007gr}.  

Single cells can interpret spatial gradients across the scale of the cell itself, either in response to external chemical gradients - such as chemotactic gradients \cite{Wadhams:2004kq} - or subcellular gradients.  Subcellular concentration gradients are found in the single cell stage of {\it C. elegans} development \cite{Griffin:2011eh}, and in single cell organisms including fission yeast \cite{Martin:2009ht, Moseley:2009hb} and bacteria \cite{Marston:1999vh, Robbins:2001bh, Thanbichler:2006do, Tsokos:2011hz, Kiekebusch:2014jg}.  In parallel with experimental work, theoretical models have demonstrated how subcellular gradients can be formed on the relevant time and spatial scales \cite{Brown:1999ue, Meyers:2006hw, Kholodenko:2006cr, Hu:2010cl, Tropini:2012ca, Howard:2012ck}.

Clustering, either of the signaling molecules or receptors, is observed in a number of subcellular signaling systems \cite{Das:2009ir, Saunders:2012jl, Veatch:2012jd, Lee:2013is}. Receptor clustering helps ensure reliable readout of input signal, for example through receptor clustering by positive feedback that enables binary on/off decisions \cite{Das:2009di}. Clustering of signaling molecules is also observed \cite{Saunders:2012jl}, potentially altering the dynamics of concentration gradient formation. 

Here, we develop a mechanistic model of cluster-mediated concentration gradient formation based on the Becker-D{\"o}ring equations of aggregation-fragmentation processes \cite{Ball:1986dk, Wattis:2006kg}. We show that such a model can produce a robust concentration gradient under (certain) biological relevant parameter conditions. In particular, we find that the process of clustering must not significantly increase the signal molecule effective lifetime. Finally, we apply the clustering model to the cortical subcellular gradient pom1p in fission yeast \cite{Padte:2006gg, Moseley:2009hb, Martin:2009ht, Saunders:2012jl}.  The modelling describes how a single component (and its interactions with itself) can create a robust subcellular concentration gradient by adapting its effective dynamical properties at different spatial positions. Theoretical approaches on the dynamics of concentration gradient formation in subcellular systems may need to be significantly different from embryonic systems.

\section{Aggregation-fragmentation model of gradient formation}

We develop a Becker-D{\"o}ring-like model of concentration gradient formation via clustering. This model takes into account aggregation, fragmentation and diffusion of clusters with only a single molecular species.  The parameters used are derived from experiments in fission yeast \cite{Saunders:2012jl}, with typical length and time scales on the order of a few microns and seconds respectively.  

\subsection{Model motivation}

We consider the Becker-D{\"o}ring equations  with conserved number and diffusion \cite{Laurencot:2002kx, Canizo:2010ik, Desvillettes:2010uh}. The motivation for such a formalism comes from studies of pom1p in fission yeast.  In time lapse movies, clusters of pom1p are not observed to coalesce (as in a Smoluchowski process \cite{Wattis:2006kg}, which describes E-Cadherin clustering \cite{TruongQuang:2013iy}) but observed to grow and decay on second time scales, apparently independently of other clusters \cite{Saunders:2012jl}.  Therefore, we assume that only monomeric molecules are taken up into a cluster and during fragmentation single molecules are released from clusters, Fig.~\ref{fig:2}A.  Cluster disassociation events from the membrane are not observed in cells \cite{Saunders:2012jl} and so we assume that disassociation involves only a single molecule at a time ({\it i.e.} when a component disassociates from both the cluster and membrane it does so without affecting the other components of the cluster), Fig.~\ref{fig:2}A. Finally, large clusters are typically not observed to join at the insertion region \cite{Saunders:2012jl} so we consider insertion only in the monomeric form. While difficult to solve analytically the existence and uniqueness of solutions of qualitatively similar models can be shown in relevant parameter regimes \cite{Canizo:2010ik}. We use a one-dimensional model of concentration gradient formation as our aim is to highlight the important general behavior of such a model.

\subsection{Model formulation}
\begin{figure}
\includegraphics{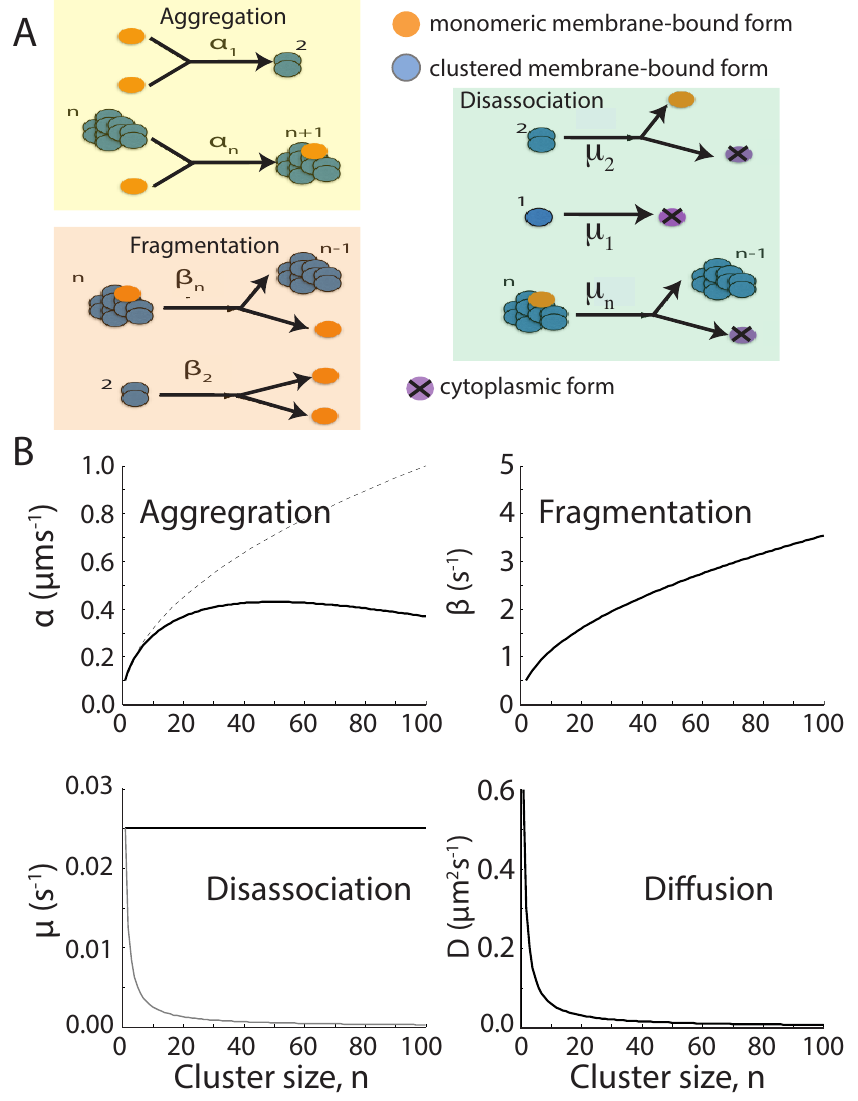}
\caption{\label{fig:2}
(Colour online) {\bf A} Reaction scheme of processes in the clustering mechanism, Eqs.~\ref{eq:main1}-\ref{eq:main3}.  Crossed circles denote particles exiting the system ({\it e.g.} through membrane disassociation). 
{\bf B} Scaling of dynamic parameters as function of cluster size.  For aggregation the dashed line corresponds to the scaling of $\alpha$ without the term $e^{-s/s_0}$.  For disassociation, the black line corresponds to equal rate of disassociation of particles in all cluster sizes. Gray line corresponds to $r_{\mu}=1$.
}
\end{figure}

We consider a mean-field discrete model of cluster formation. The reaction schemes are shown in Fig.~\ref{fig:2}A and result in the following equations for clusters where $n_s$ represents the concentration of a cluster containing $s$ molecules:
\begin{eqnarray}
\label{eq:main1}
\nonumber\frac{\partial{n}_1}{\partial{t}} =&D_1&\frac{\partial^2n_1}{\partial{x}^2}-\mu_1n_1+\mu_2n_2 \\ 
 &-&\sum_{s=2}^{s_{max}}j_s-j_2+J\delta{(x)} \\
\label{eq:main2}
\nonumber\frac{\partial{n}_s}{\partial{t}} =&D_s&\frac{\partial^2n_s}{\partial{x}^2} + (j_s - \mu_ssn_s) \\
 &-&(j_{s+1}-\mu_{s+1}(s+1)n_{s+1})\text{      for  } s>1 \\
\label{eq:main3}
j_s = &\alpha_{s-1}&n_1n_{s-1}-\beta_sn_s \,.
\end{eqnarray}
$D_s$ denotes the diffusion coefficient, $\mu_s$ the membrane disassociation rate and $\alpha_s$, $\beta_s$ denote the aggregation and fragmentation rates respectively for a cluster containing $s$ molecules.  $J$ is the monomeric insertion rate and boundary conditions are $\frac{\partial{n_s}}{\partial{x}}|_{x=0,L} = 0$ for $s\ge2$.

\subsection{Parameter scaling}

Eqs.~\ref{eq:main1}-\ref{eq:main3} have a large parameter space, increasing with $s_{max}$, the maximum cluster size. However, using biophysical arguments we can reduce the parameter space to seven non-dimensional parameters as discussed below.  

{\it Diffusion}: We take $D_s = D_1s^{-r_D}$  where $r_D = 1$, Fig.~\ref{fig:2}B, consistent with experiments on clustered protein membrane diffusion \cite{Gambin:2006jd,Saunders:2012jl} (note, this contrasts with theoretical predictions of logarithmic diffusion scaling on biological membranes \cite{Saffman:1975fq}). We have confirmed our main conclusions hold for $r_D=2/3$ (not shown).

{\it Disassociation}: Experimentally, large clusters are not observed to disassociate from the membrane\cite{Saunders:2012jl}.  Therefore, we assume only single molecules disassociate in each reaction.  We take $\mu_s=\mu_1s^{-r_{\mu}}$. $r_{\mu}=0$ if clustering has no membrane-stabilizing effect ({\it i.e.} the rate of monomer disassociation is equivalent from all clusters). We also consider the case $r_{\mu}=1$, {\it i.e.} clustering stabilizes the signaling molecule within the cluster, Fig.~\ref{fig:2}B. We shall see that the latter scenario results in a non-robust gradient and is a key result of this analysis.

{\it Aggregation}: Aggregation depends on the cluster size - bigger clusters are more likely to collide with and aggregate a monomer.  How exactly aggregation depends on the cluster topology is unclear and here we consider $\alpha_s = \alpha_1s^{r_{\alpha}}$ with $r_{\alpha}=1/2$. Experimentally, there is an upper limit on the cluster size \cite{Saunders:2012jl}, so an additional term within the aggregation parameter is included to limit the maximum size of the clusters: $\alpha_s=\alpha_1s^{r_{\alpha}}e^{-s/s_0}$ where $s_0=100$ \cite{TruongQuang:2013iy}, Fig.~\ref{fig:2}B. Such limitation on the maximum cluster size could also have been implemented by including a component in the fragmentation rate (below).

{\it Fragmentation}: Scaling of the fragmentation rate is likely to occur via a similar mechanism to aggregation. Hence, we consider $\beta_s=\beta_1s^{r_\beta}$ with $r_{\beta}=r_{\alpha}=1/2$. We have confirmed that using $r_{\beta,\alpha}=1/3$ or $r_{\beta,\alpha}=2/3$ does not significantly alter our key conclusions so long as $r_{\beta}=r_{\alpha}$. We choose the values of $\alpha_1$ and $\beta_1$ such that cluster dynamics are on the order of a few seconds, consistent with pom1p cluster dynamics \cite{Saunders:2012jl}. Note, we use $\beta_1$ as the lowest order for fragmentation (even though only terms $\beta_s$, $s\ge2$ appear in Eqs.~\ref{eq:main1}-\ref{eq:main3}) for clarity in representing the scaling.

Non-dimensionalizing by substituing $t=\tau/\mu_1$, $\rho_s = \frac{J}{\sqrt{D_1\mu_1}}\phi_s$ and $x=u\sqrt{D_1/\mu_1}$, along with the scaling arguments, reduces Eqs.~\ref{eq:main1}-\ref{eq:main3} to
\begin{eqnarray}
\nonumber\frac{\partial\phi_1}{\partial\tau} &=& \frac{\partial^2\phi_1}{\partial{u}^2}  - \sum_{s=2}^{s_{max}}\left(\kappa\phi_1s^{r_{\alpha}}e^{-s/s_0}-\tilde{\beta}s^{r_{\beta}}\right)\phi_s \\
 &\text{ }&- \phi_1+ \left[2^{1-r_{\mu}}\phi_2 + \tilde{\beta}\phi_2-\kappa\phi_1^2\right]+\delta(u) \\
\nonumber \frac{\partial\phi_s}{\partial\tau} &=& s^{-r_D}\frac{\partial^2\phi_s}{\partial{u}^2}  - s^{1-r_{\mu}}\phi_s+(s+1)^{1-r_{\mu}}\phi_{s+1} \\
\nonumber  &\text{ }&-\kappa\phi_1e^{-s/s_0}\left(s^r_{\alpha}\phi_s-(s-1)^{r_{\alpha}}e^{1/s_0}\phi_{s-1}\right) \\
 &\text{ }&-\tilde{\beta}\left(s^{r_{\beta}}\phi_s - (s+1)^{r_{\beta}}\phi_{s+1}\right)\,,
\end{eqnarray}
where $\kappa = J\alpha_1/\sqrt{D\mu_1^3}$ and $\tilde{\beta} = \beta_1/\mu_1$. The large parameter space in Eqs.~\ref{eq:main1}- \ref{eq:main3} has been reduced to seven dimensionless parameters (Table~\ref{tab:table2}), independent of $s_{max}$. Four of these seven parameters, $r_{\alpha,\beta,D,\mu}$ are constrained by physical arguments as described above.  The phenomenological cluster size factor $s_0$ is limited by experimental observation of cluster sizes \cite{Saunders:2012jl}.  Only $\kappa$ and $\tilde{\beta}$ are free parameters and hence, despite the apparent complexity of Eqs.~\ref{eq:main1}- \ref{eq:main3}, the dynamic behavior of the system is effectively described by just two parameters.

\begin{table}[b]
\caption{\label{tab:table2}%
Dimensionless parameters in clustering model}
\begin{ruledtabular}
\begin{tabular}{ccc}
{\textrm{Parameter}}&{\textrm{Value}}&{\textrm{Note}}\\
\colrule
$r_{\alpha}$, $s_0$ & $\frac{1}{2}$, 100 & $s_0 = 100$ constrain  \\
 & & maximum cluster size \\
$r_{\beta}$ & $\frac{1}{2}$ & $r_{\beta} = r_{\alpha}$ \\
$r_{\mu}$ & 0, 1 & $r_{\mu}>0$ results in stabilization\\
 & & of monomers in larger clusters \\
$r_{D}$ & 1 & See \cite{Saunders:2012jl,Gambin:2006jd} \\
$\kappa=\frac{J\alpha_1}{\sqrt{D_1\mu_1^3}}$ & $\sim425$ & Defines range of clustering effects \\
$\tilde{\beta}$ & $\sim20$ & Fragmentation occurs on shorter\\
 & & timescales than disassociation \\
\end{tabular}
\end{ruledtabular}
\end{table}

\begin{figure}
\includegraphics{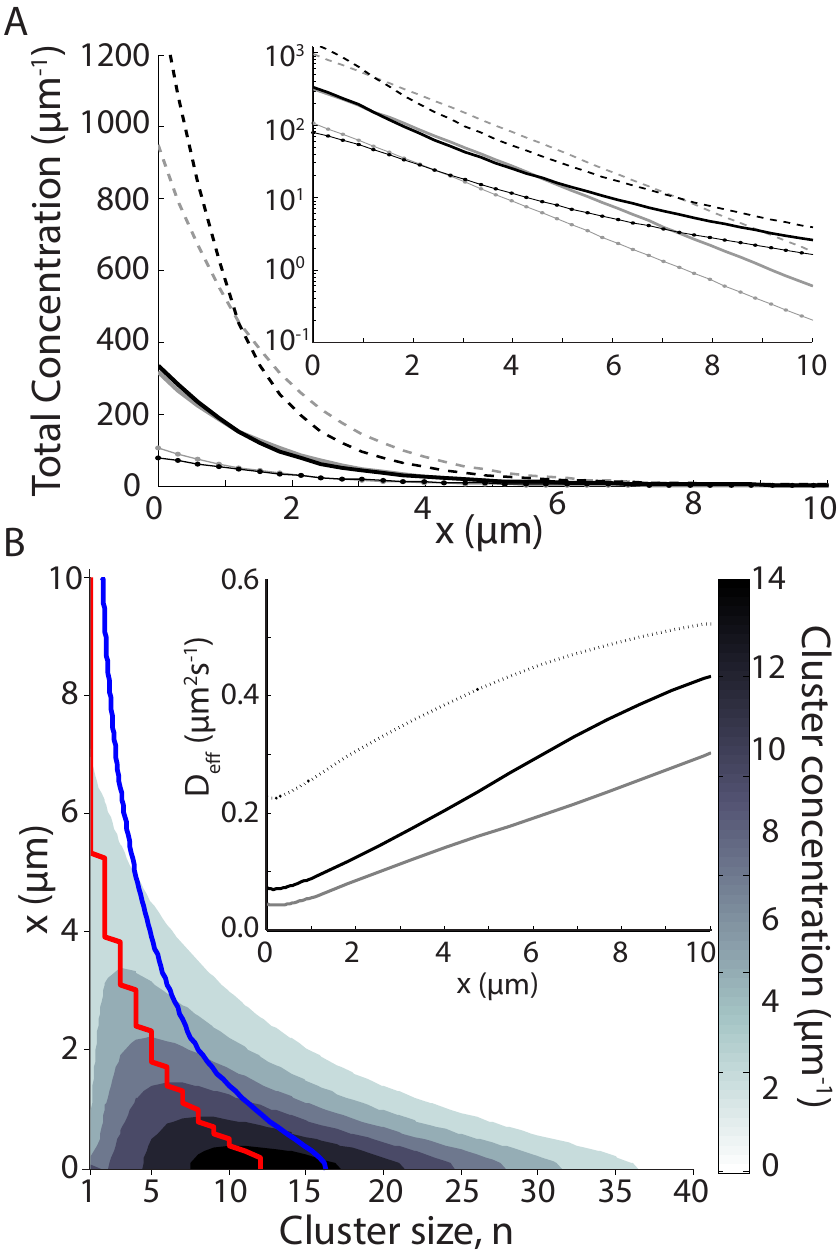}
\caption{\label{fig:3}
(Colour online) {\bf A} Concentration profile, Eq.~\ref{eq:profile} with $\kappa=425$ and $\tilde{\beta}=20$. Dotted and dashed black lines correspond to $\kappa\approx142$  ($J\rightarrow{J}/3$) and $\kappa =1275$  ($J\rightarrow3J$) respectively. Gray lines show solutions of a model without clustering with same insertion and disassociation rates and $D=0.1\mu{m}s^{-1}$ (see SDD model in Table~\ref{tab:table1}).  Inset shows profiles on logarithmic scale. {\bf B} Cluster concentration as function of position, with darker color representing higher concentration.  Red (step-like) and blue (smooth) lines correspond to the cluster size with highest concentration and mean cluster concentration respectively. Inset: effective diffusion, Eq.~\ref{eq:diff}, as function of position for $\tilde{\beta}=20$  (black),  $\tilde{\beta}=200$ (dashed) and $\tilde{\beta}=2$ (gray). Parameters as Fig.~\ref{fig:2}B unless otherwise stated.}
\end{figure}

\subsection{Concentration gradient profile}
We solve the above system of equations (up to $s_{max}$ cluster size, typically 200) using Matlab pde45 and confirm the steady-state distribution using the Matlab ode solver bvp4c.  The total concentration profile, defined as
\begin{equation}
\label{eq:profile}
N_T = \sum_{s=1}^{s_{max}}sn_s\,,
\end{equation}
is shown in steady-state in Fig.~\ref{fig:3}A, for a range of $\kappa$. Unsurprisingly, large clusters are localized to the source region, Fig.~\ref{fig:3}B.  The effective diffusion,
\begin{equation}
\label{eq:diff}
D_{eff}(x) = \left(\sum_{s=1}^{s_{max}}sn_sD_s\right)/N_T \,,
\end{equation}
is a function of position, increasing away from the source, Fig.~\ref{fig:3}B inset. The timescales of clustering [42] and protein dynamics are consistent with observations of pom1p {\it in vivo} \cite{Saunders:2012jl}.

What biophysical processes do $\kappa$ and $\tilde{\beta}$ represent? $\tilde{\beta}$ effectively defines the relative lifetime of molecules in clusters before disassociation, which in turn alters the effective diffusion, Fig.~\ref{fig:3}B inset.  $\kappa$ compares the effects of insertion and aggregation (increasing either amplifies clustering) with diffusion and disassociation (increasing either reduces clustering).  Small $\kappa$ corresponds to weak clustering and the diffusive dynamics of the monomer dominate. Large $\kappa$ results in clustering dominating the dynamics with a resulting steep concentration gradient.  Using biologically plausible parameter values \cite{Saunders:2012jl}, both $\kappa$ and $\tilde{\beta}$ occur at values that allow large clusters to form (to give benefits of modulating diffusion) without permitting very large clusters to dominate the dynamics, particularly away from the source region.

\section{Robustness of cluster-mediated concentration gradient formation}

We have demonstrated that a clustering model can produce a concentration gradient on similar spatial and temporal scales observed in subcellular systems. Ultimately, the concentration profile must be able to impart precise ({\it i.e. robust}) information to the cell. Below, we explore whether such concentration gradients can be robust to relevant biological fluctuations.  To gain a qualitative understanding, we first discuss a phenomenological model incorporating concentration-dependent diffusion before discussing the effects of noise on the clustering model.

\subsection{Concentration-dependent diffusion}
We consider a one-dimensional reaction-diffusion equation with concentration-dependent diffusion (CDD), $D(\rho)$, a function of the local protein concentration, $\rho$:  
\begin{equation}
\label{eq:phen}
\frac{\partial\rho}{\partial{t}}=\frac{\partial}{\partial{x}}\left(D(\rho)\frac{\partial\rho}{\partial{x}}\right)-\mu\rho \,,
\end{equation}
with boundary conditions $D(\rho(0))\frac{\partial\rho}{\partial{x}}|_{x=0}+J=0$ where $J$ is the protein insertion rate and $\rho(x\rightarrow\infty)=0$.  We consider the case $D(\rho)=D_0(\bar{\rho}/\rho)^r$.  The solution to Eq.~\ref{eq:phen} then has steady-state solution $\rho(x)=B(x+x_1)^{-m}$ where $m=2/r$, $B = \lambda^m\bar{\rho}[m(m-1)]^{m/2}$ and $x_1 = \lambda\left(\frac{\bar{\rho}}{\rho_0^S}\right)^{\frac{1}{m-1}}g(m)$, where $g(m) = \left(m^{\frac{m}{m-2}}(m-1)\right)^{\frac{m-2}{2(m-1)}}$, $\rho^S_0=J/\sqrt{D_0\mu}$ and $\lambda=\sqrt{D_0/\mu}$, Table~\ref{tab:table1}.

\begin{table}[b]
\caption{\label{tab:table1}
Steady state solutions to SDD, NLD and CDD models. For NLD model solution for non-linear degradation term $-\alpha\rho^2$ is shown \cite{Eldar:2003eh}. CDD model solution for $r=1/2$.}
\begin{ruledtabular}
\begin{tabular}{cccc}
Model&$\rho(x,t\rightarrow\infty)$&Parameters \\
\colrule
SDD & $\frac{J}{\sqrt{D_0\mu_0}}e^{-x/\lambda}$ & $\lambda = \sqrt{\frac{D_0}{\mu_0}}\approx1.6{\mu}m$ \\
NLD & $\frac{A}{(x+x_0)^2}$& $A=6D_0/\alpha$, $x_0=\left(\frac{12D_0^2}{J\alpha}\right)^{1/3}$ \\
CDD & $\frac{B}{(x+x_1)^4}$& $B=144\bar{\rho}\lambda^4$, $x_1=\lambda\left(\frac{48\bar{\rho}}{\rho_0^s}\right)^{1/3}$ \\
\end{tabular}
\end{ruledtabular}
\end{table}
\begin{figure}
\includegraphics{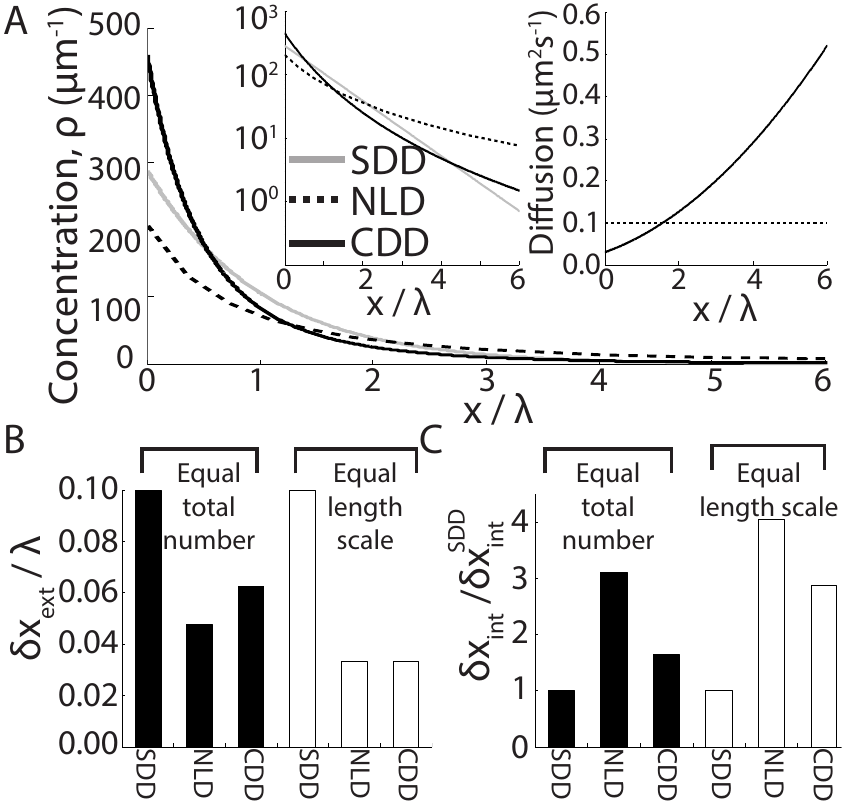}
\caption{\label{fig:1}
{\bf A} Steady-state profiles for SDD (gray), NLD (dashed) and CDD (black) models (Table~\ref{tab:table1}). SDD model: $J = 16.7s^{-1}$, $D_0 = 0.1\mu{m}^2s^{-1}$ and $\mu=0.03s^{-1}$.  $\bar{\rho}=40{\mu}m^{-1}$ and non-linear degradation rate $\alpha$ were chosen such that the total particle number was equal in all models with same $J$, $D_0$ and $\mu_0$ and such that the average diffusion in the CDD model for $0<x/\lambda<3$ equals $D_0$.  Left inset: same as main panel but on y-log scale. Right inset: $D(\rho)$ in CDD model, dashed line corresponds to $D=0.1\mu{m}^2s^{-1}$.
{\bf B} Error in defining spatial position, Eq.~\ref{eq:dx}, due to variations in $J$, given by $\delta{J}= 0.1J$. Black bars denote parameters as {\bf A}. Clear bars correspond to equal characteristic length scale, $\lambda=x_0=x_1$ ($\bar{\rho}=6{\mu}m^{-1}$).  Smaller values denote more robust profiles.
{\bf C} Robustness of the NLD and CDD models due to intrinsic fluctuations at $x/\lambda=2$ relative to SDD model. Black and clear bar notation same as {\bf B}. Though the specific fractions vary as a function of position, the trend of $\delta{x}_{int}^{SDD}<\delta{x}_{int}^{CDD}<\delta{x}_{int}^{NLD}$ typically holds in the region $0<x<3\lambda$.}
\end{figure}
In Fig.~\ref{fig:1}A we compare steady-state profiles for the NLD and CDD models (right inset shows the behavior of the diffusion coefficient in the CDD model as a function of position) with the scenario of linear diffusion ($D(\rho)=D_0$ in Eq.~\ref{eq:phen}) and degradation (SDD model), Table~\ref{tab:table1}.  We define robustness as the positional error, $\delta{x}$ in defining a boundary at a threshold concentration \cite{Eldar:2003eh}
\begin{equation}
\label{eq:dx}
\delta{x}(x,t)=\delta\rho(x,t) / |\rho(x,t)'|
\end{equation}
due to concentration fluctuations $\delta\rho(x,t)$ ($\rho'$ is the spatial derivative) from, for example, variations in $J$.  $\delta{x}$ is dependent on when and where measurement occurs. We focus on the spatial position in steady-state for distances around $2-3\mu{m}$ from the source, consistent with boundaries in fission yeast. Interestingly, in steady-state, $\delta{x}$ is independent of position for all three models if the concentration fluctuations are due to variation in the injection rate $J$, Fig. 3B, \cite{Eldar:2003eh}. Previously, it has been shown that a model with non-linear degradation (NLD) can produce concentration profiles that are robust to variations in the insertion rate \cite{Eldar:2003eh}. In Fig.~\ref{fig:1}B we demonstrate that the CDD model, just as with the NLD model, is more robust to variations in the insertion rate compared to the SDD model when $0<r<1$. 

Another source of error are stochastic biochemical (intrinsic) fluctuations, relevant in both embryonic and subcellular systems \cite{He:2010jt, Saunders:2012jl, Liu:2013gv, Bothma:2014jq}.  Such fluctuations are typically well-described by Poisson statistics \cite{Tostevin:2007gl,Saunders:2009iy}:
\begin{equation}
\label{eq:int_time_ave}
\delta\rho_{int} = a \sqrt{\frac{\rho}{DT}}\,,
\end{equation}
where $a$ is a constant that is assumed to be model independent and $T$ is the averaging period.  NLD models are generally less robust to intrinsic fluctuations \cite{Saunders:2009iy}.  However, in the CDD model the diffusion coefficient increases with distance from the source, which in turn increases the effects of time averaging \cite{Tostevin:2007gl} and hence reduces the detrimental effects of intrinsic fluctuations.  In Fig.~\ref{fig:1}C we show that the SDD model is most robust to such variations at $x/\lambda=2$, but concentration-dependent diffusion results in more robust gradients than those formed by non-linear degradation processes.

\subsection{Robustness of clustering model}
Having developed a qualitative understanding of the robustness of concentration-dependent diffusion to relevant fluctuations we now discuss the robustness of the full clustering model, Eqs~\ref{eq:main1}-\ref{eq:main3}.

\begin{figure}
\includegraphics{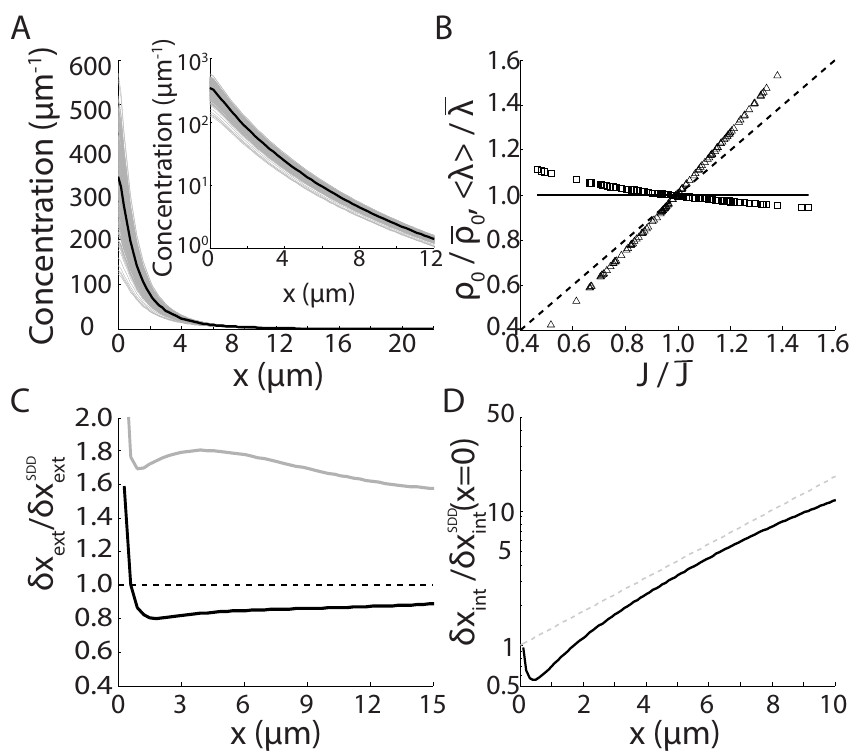}
\caption{\label{fig:4}
{\bf A} 250 profiles (gray) generated from parameters shown in Fig.~\ref{fig:2}B, but with 20\% fluctuations in insertion rate. Black line is mean concentration. Inset is same with y-axis on logarithmic scale.
{\bf B} Scaling of concentration at $x=0$ (triangles for clustering model, dashed line for SDD model) and decay length $\langle\lambda\rangle_{2-3\mu{m}}$ (squares for clustering model, line for SDD model), see text for definition, as function of insertion rate normalized by their value for mean insertion rate, denoted by $\bar{J}$, $\bar{\rho_0}$ and $\bar\lambda$.
{\bf C} Positional error of the clustering model relative to the SDD model with equivalent kinetic parameters. Black line corresponds to clustering model with equal disassociation rate in all clusters, $r_{\mu}=0$. Gray line corresponds to clustering model with $r_{\mu}=1$. 
{\bf D} The positional accuracy of the clustering model in presence of intrinsic fluctuations, Eq.~\ref{eq:int_time_ave}, normalized by the equivalent accuracy of the SDD model at $x=0$. Equivalent result for SDD model shown as gray line.
}
\end{figure}

To test the robustness of the clustering model to variations in protein insertion we first created 250 profiles using parameters in Fig.~\ref{fig:2} but with the insertion rate Gaussian distributed (mean $\bar{J}$, standard deviation $\delta{J}=0.2\bar{J}$), Fig.~\ref{fig:4}A. Fig.~\ref{fig:4}B shows that the concentration at $x=0$ (normalized by the value when $J=\bar{J}$) increases rapidly (faster than equivalent SDD model) as $J$ increases; as expected, the fluctuations at the source are increased in the clustering model.  Following \cite{Eldar:2003eh}, we define the characteristic length scale of the profile as 
\begin{equation}
\lambda_{eff}(x)=\rho(x)/|\rho'(x)| \,.
\end{equation}  
$\lambda_{eff}$ is a function of position and henceforth we consider the mean $\lambda_{eff}$ in the range $2\mu{m}<x<3\mu{m}$, $\langle\lambda\rangle_{2-3\mu{m}}$. In Fig.~\ref{fig:4}B, we see that $\langle\lambda\rangle_{2-3\mu{m}}$ scales inversely with increasing $J$ (unlike the SDD model). Therefore, the increases variation at $x=0$ is compensated for by adaptation in the characteristic profile length.  This results in the cluster model providing  precise spatial information, Fig.~\ref{fig:4}C, compared to SDD model. In Fig.~\ref{fig:4}C we also show that the positional accuracy of the clustering model is significantly reduced when clustering stabilizes the protein. Therefore, the clustering model is only robust if individual monomers only spend a relatively short period in each cluster - if they are too stable then the robustness is lost.

The trend of $\lambda_{eff}$ being inversely proportional to $J$ can be derived using a two-state (monomer / clustered) model \cite{Saunders:2012jl}. However, in such a model fluctuations at the source increase especially quickly with increasing $\delta{J}$.  Here, our more precise analysis shows that clustering results in a robust gradient when comparing the competing effects of $\rho(x=0)$ and $\lambda_{eff}$ fluctuations with changing $J$ (note that the particular effect is position dependent).  Furthermore, we see that the robustness is due to the effective spatial adaptation of the diffusion coefficient, which is highly dependent on $\tilde{\beta}$, Fig.~\ref{fig:3}B, as well as on the scaling of the disassociation rate with cluster size. The simplifications used in the two-state model meant that the critical relationship between fragmentation and disassociation rates was not appreciated, highlighting a further advantage of our more detailed approach.

Finally, we compare the effects of intrinsic fluctuations on the robustness of the clustering model.  The processes of clustering are non-linear and therefore the Poisson approximation is less valid.  However, direct experimental measurement of intrinsic fluctuations in a subcellular gradient \cite{Saunders:2012jl} suggest that away from the source region ($x\gtrsim2\mu{m}$), where diffusion is the predominant dynamic mechanism, that intrinsic fluctuations are approximated closely by Poisson statistics. Therefore, Eq.~\ref{eq:int_time_ave} is a reasonable approximation to the intrinsic noise here, though likely represents a lower bound on the true intrinsic noise due to neglected non-linear effects from clustering processes. In Fig.~\ref{fig:4}D we show the positional accuracy of the cluster model given Poisson distributed intrinsic fluctuations, compared to the SDD model with similar profile shape and protein disassociation rate. The two models have qualitatively similar sensitivity to intrinsic fluctuations due to their similar profile shape, Eq.~\ref{eq:int_time_ave}.  The clustering model is less sensitive to intrinsic fluctuations than the CDD and NLD models discussed above as the latter two models have algebraic, rather than exponential-like, profiles \cite{Saunders:2009iy}.

\section{Pom1p subcellular gradient}
We apply our clustering model to the specific case of pom1p in fission yeast and its repression of the downstream target cdr2p \cite{Martin:2009ht, Moseley:2009hb, Rincon:2014hq}.  We incorporate spatially distributed pom1p insertion in the polar region \cite{Saunders:2012jl, Howard:2012ck}.  The formation of the pom1p subcellular gradient has been modelled previously \cite{Vilela:2010br,Tostevin:2011jo,Saunders:2012jl}.  However, these approaches either did not consider clustering \cite{Vilela:2010br,Tostevin:2011jo} or presented only a qualitative model of clustering with only two states \cite{Saunders:2012jl}.  

\begin{figure}
\includegraphics{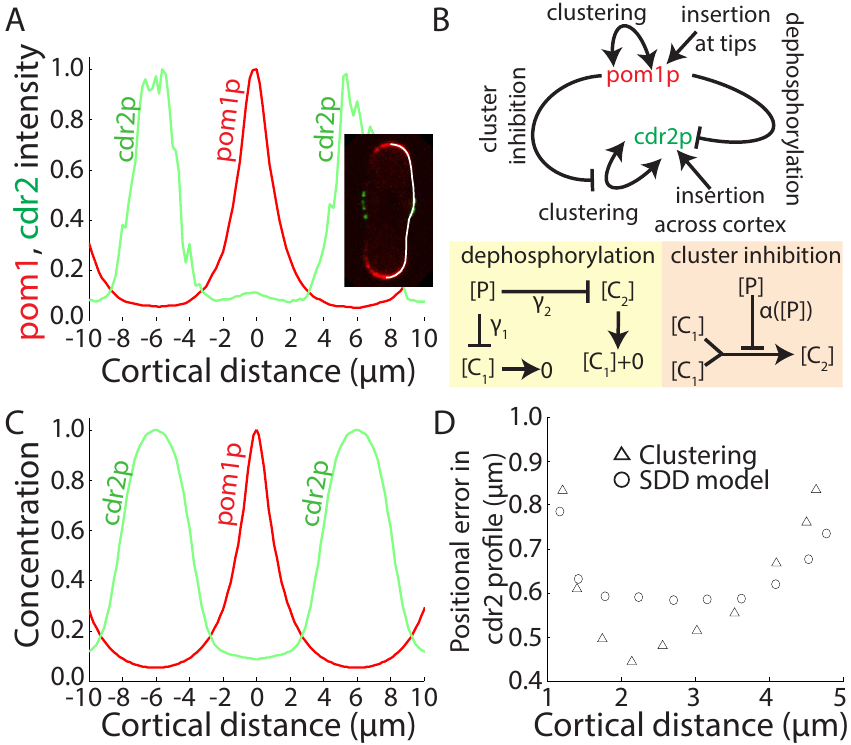}
\caption{\label{fig:5}
(Colour online) {\bf A} Experimentally measured (normalized) intensity profiles for pom1p (red) and cdr2 (green) on the cortex of fission yeast cells ($n=14$), see \cite{Pan:2014ks} for details. Distance defined as the distance around the cortex of the cell starting from the tip, as shown by the white line in inset. Inset: Midplane confocal image of fission yeast expressing pom1p-tomato in the tip (red) and cdr2-GFP at the cell center (green).
{\bf B} Interactions between pom1p and cdr2p on the membrane. Pom1p actively disassociates cdr2p from the membrane (yellow box) and also inhibits the ability of cdr2p to cluster (orange box). 
{\bf C} Normalized concentration profiles for pom1p (red) and cdr2p (green) generated from Eqs.~\ref{eq:main1}-\ref{eq:main3}, \ref{eq:cdr21},\ref{eq:cdr22}.  Parameters for pom1p as Fig.~\ref{fig:2}. For cdr2p, $\eta_1=\eta_2/2=10^{-3}s^{-1}$, $J_c=2.6s^{-1}$, $[P]_0=20\mu{m}^{-1}$, $D_{c,1} = 0.5{\mu}m^2s^{-1}$, $D_{c,2} = 0.01{\mu}m^2s^{-1}$, $\mu_{c,1}=\mu_{c,2}/2=2.5\times10^{-3}s^{-1}$, $\alpha_c=0.1\alpha_1$ and $\beta_c=0.1\beta_1$.
{\bf D} Positional error in the cdr2p profile when the pom1p insertion rate has 25\% Gaussian variation.  Results for clustering model (triangles) and equivalent SDD model (circles) shown.
}
\end{figure}

Cdr2p itself is known to cluster; indeed, it forms significantly larger clusters with around 80-100 molecules in the clusters localized to the cell center \cite{Pan:2014ks}. Pom1p has a dual effect on cdr2p. First, it is involved in cdr2p dephosphorylation, resulting in cdr2p membrane disassociation.  Second, it inhibits cdr2p cluster formation.  This double mechanism of repression helps produce the sharp response of cdr2p to pom1p inhibition \cite{Rincon:2014hq}, Fig.~\ref{fig:5}A,B.  We use a phenomenological model of cdr2p cluster formation since our focus is on pom1p and how accurately it can define the cdr2p boundary. Therefore, we use a two state model for cdr2p (monomeric or clustered, similar to \cite{Saunders:2012jl}) with additional interactions between pom1p (where $[P]$ denotes the total pom1p concentration at a particular position, regardless of the particular cluster distribution, Eq.~\ref{eq:profile}) and cdr2p (where $[C_{1,2}]$ denotes cdr2p concentration in monomeric and clustered forms respectively), Fig.~\ref{fig:5}B.  Cdr2p is assumed to be inserted uniformly (rate $J_c$) across the membrane \cite{Pan:2014ks}.
\begin{eqnarray}
\nonumber \frac{\partial[C_1]}{\partial{t}} = &D_{c,1}&\frac{\partial^2[C_1]}{\partial{x}^2}-\mu_{c,1}[C_1]-2\alpha_c([P])[C_1]^2+2\beta_c[C_2]\\
\label{eq:cdr21} &+&\mu_{c,2}[C_2]-[P](\gamma_1[C_1]-\gamma_2[C_2]) + J_c \\
\nonumber \frac{\partial[C_2]}{\partial{t}} = &D_{c,2}&\frac{\partial^2[C_2]}{\partial{x}^2} -\mu_{c,2}[C_2]+\alpha_c([P])[C_1]^2\\
\label{eq:cdr22} &-&\beta_c[C_2] - \gamma_2[P][C_2] 
\end{eqnarray}
The direct inhibition of cdr2p on the cortex by pom1p is approximated by the term $-\gamma_{1,2}[C_{1,2}][P]$ and the aggregation factor $\alpha_c$ is now dependent on the concentration of pom1p, Fig.~\ref{fig:5}B. We assume a Hill-like behavior: $\alpha_c([P]) = \alpha_{c,0}(1+([P]/[P]_0)^4)^{-1}$. This simplified model recapitulates the observed pom1p and cdr2p profiles, Fig.~\ref{fig:5}C. The fitting from the above model to the measured pom1p and cdr2p profiles represents a significant improvement over previous models \cite{Vilela:2010br,Tostevin:2011jo,Saunders:2012jl} both in replicating the spatial profiles and reproducing the dynamics of clustering and gradient formation. 

To test the system robustness we created 200 pom1p profiles, with insertion rate normally distributed with standard deviation 25\% of the mean and subsequent cdr2p profiles using Eqs.~\ref{eq:cdr21}-\ref{eq:cdr22}.  At each position investigated we found the mean total cdr2p concentration and used this to define the threshold concentration for that position. For each individual cdr2p profile we then measured the position where it had each particular threshold concentration and hence calculated the positional precision of cdr2p specification by calculating the standard deviation in these positions.  Near the source there is large variation due to the big intensity changes in pom1p between cells. Near the cell center there is large error due to the pom1p profile becoming increasingly flat.  However, around $2-3\mu{m}$ from the source which corresponds to the region where the boundary between pom1p and cdr2p is defined - we see that the cdr2p can be positioned more accurately (compared with equivalent SDD model) by pom1p when it clusters, Fig.~\ref{fig:5}D. Of course, including intrinsic fluctuations would decrease the positional precision so the given accuracy represents a best case scenario (experimentally, errors of around $1\mu{m}$ are typically observed \cite{Saunders:2012jl}).  In conclusion, a dynamic clustering mechanism for concentration gradient formation can provide robust positional information on relevant time and spatial scales for a biologically plausible scenario.

\section{Discussion and conclusions}

Previous modelling of clustering within cells has predominantly analyzed receptor clustering.  Here, we have focused on the role of clustering in the formation of concentration gradients and demonstrated that subcellular gradients can be formed via clustering on realistic spatial and temporal scales. This work represents a significant advance on previous models of subcellular concentration gradient formation \cite{Tostevin:2007gl, Tostevin:2011jo, Jilkine:2011fs, Saunders:2012jl} as it accounts for protein clustering and diffusion in a mechanistic (though still relatively straightforward) framework that also incorporates realistic protein dynamics, and allows predictions to be made about the behaviour of specific dynamic components (see below).

Our modeling enables the following predictions regarding the clustering of signaling molecules. (i) the process of clustering does not significantly stabilize (whether that be by extending protein lifetime or membrane association time) the individual molecules within the concentration gradient.  If clustering significantly stabilizes the protein in the relevant system then the resultant chemical gradient is not robust.  This result is qualitatively consistent with the dynamics of pom1p, where the cluster lifetime is significantly shorter than the pom1p lifetime on the membrane \cite{Saunders:2012jl}.  (ii) The ratio of the fragmentation rate to disassociation rate plays an important role in the gradient formation.  Either too small or big a ratio results in reduced spatial diffusion modulation and hence less robust concentration gradients. Therefore, systems that use clustering in concentration gradient formation are likely to have carefully tuned fragmentation and disassociation rates and  experimental perturbation of either should result in significantly reduced robustness.  (iii) Clustering is favorable in systems that have a single decision to make ({\it e.g.} placement of division boundary) but it is less likely to be used in systems that specify multiple boundaries. Near the source there is increased inaccuracy due to larger fluctuations in the concentration and at very large distances the profile becomes very flat due to only small, fast, clusters being present.  Depending on the specific parameters, there will likely be an optimal region for concentration gradient interpretation. 

Given the advantages described above, why is clustering of signaling molecules ({\it i.e.} morphogens) not observed more commonly in multicellular systems? In single cells where a simple decision is made by the signaling pathway ({\it e.g.} where to define the cell center) then clustering may be advantageous as dynamic parameters can be tuned to maximize precision at the relevant position but for morphogen gradients, that typically define three or more threshold positions across their spatial range, clustering may not be beneficial. Further, multicellular organisms typically have more time and complexity to adjust for variation in the input signal, such as via feedback networks \cite{Manu:2009fl, HaskelIttah:2012ft}.  We note that the Hedgehog signaling protein is observed to cluster \cite{Gradilla:2013ez} but this is likely due to its need for a chaperone to aid it in traversing through the intercellular space due to its hydrophobic nature \cite{Gradilla:2013ez}.

We have considered a one-dimensional mean-field scenario.  Two-dimensional simulations of the clustering model could be interesting, as the stochastic noise in such a (plausible) scenario is non-trivial, particularly near the insertion region. However, experimental evidence suggests that the approximations used here are relevant in the signaling region ({\it i.e.} away from the source).  Further, extension to Smoluchowski processes where clusters directly interact with each other may be interesting but, as noted above, such a scenario is not consistent with current experimental observations of subcellular concentration gradients.

Overall, we have presented a quantitative framework for understanding subcellular concentration gradient formation.  In particular, our model can simultaneously replicate observed experimental spatial profiles and dynamics. Importantly, despite the apparent complexity of such a clustering model, through biophysical arguments we reduced our model to two relevant parameters to describe the concentration gradient.  The resulting concentration profile is robust to relevant biochemical fluctuations so long as the process of clustering does not significantly stabilize the signaling molecule. 

\begin{acknowledgments}
We thank Martin Howard, Lars Hufnagel and Jacques Prost for discussions on the modeling and Fred Chang for discussions regarding clustering experiments on pom1p. We are grateful to Celine Stoecklin for work investigating the sensitivity of the clustering model to changes in the scaling parameters.  Fred Chang and Ignacio Flor Parra kindly provided the experimental data in Fig.~\ref{fig:5}.  We thank Nils Gaulthier, Christopher Amourda and Virgile Viasnoff for helpful comments on the manuscript.  This work was funded by the National Research Foundation, Singapore.\end{acknowledgments}

\bibliography{Saunders_Submission}
\end{document}